\documentclass[aps,physrev,onecolumn,superscriptaddress,10pt]{revtex4-2}

\usepackage[T1]{fontenc}
\usepackage{graphicx}
\usepackage{multirow}
\usepackage{makecell}
\usepackage{hyperref}

\begin{document}

\title{Through-bottle spectroscopy as a tool for quality control and anti-counterfeiting of Brandy and Cognac}

\newcommand{\StA}{SUPA, School of Physics and Astronomy, University of St Andrews, North Haugh, St Andrews KY16 9SS, United Kingdom}
\newcommand{\Henn}{Jas Hennessy \& Co, Rue de la Richonne, 16100 Cognac, France}

\author{George O. Dwapanyin}
\affiliation{\StA}
\author{Edward I. Appleton}
\affiliation{\StA}
\author{Stella Corsetti}
\affiliation{\StA}
\author{Charles Descoins}
\affiliation{\Henn}
\author{Xavier Poitou}
\affiliation{\Henn}
\author{Kishan Dholakia}
\affiliation{\StA}
\affiliation{School of Biological Sciences, The University of Adelaide, Adelaide, Australia}
\affiliation{Centre of Light for Life, The University of Adelaide, Adelaide, Australia}
\author{Graham D. Bruce}
\email{gdb2@st-andrews.ac.uk}
\affiliation{\StA}

\begin{abstract}
Counterfeiting of premium spirits poses significant economic and health risks, that could be tackled by robust, accurate, portable and non-destructive through-bottle measurements.  Here, we demonstrate the capability of focus-matched inverse spatially offset spectroscopy, combining fluorescence and Raman signals, for authenticating Cognac and Brandy. The technique accurately identifies age classification, bottling year, spoilage due to elevated storage temperatures, and distinguishes between Cognac brands. Critically, our method effectively differentiates genuine Cognacs from counterfeit products, correctly identifying 98\% of counterfeit samples. This shows the promise of through-bottle spectroscopy as a powerful tool for supply chain integrity and consumer protection in the high-value spirits market.
\end{abstract}

\maketitle

\section*{Introduction}

The allure of high-profit margins associated with premium spirits makes them prime targets for counterfeiters, leading to economic loss and to the sale of products harmful to human health. The counterfeit market for wine, beer, and spirits has been estimated to cause losses of €3 billion annually within the EU \cite{FAP}. Traditional laboratory techniques, such as mass spectrometry, nuclear magnetic resonance spectroscopy, gas chromatography, and liquid chromatography, are employed to ensure the quality, safety, and authenticity of spirits \cite{power2020brief}. However, these methods are often costly, time-consuming, and require sample removal, limiting their applicability for online monitoring and in-field analysis. This necessitates the development of portable, non-destructive analytical techniques for rapid and accurate authentication. Through-bottle sensing is particularly desirable in the area of anti-counterfeiting, as it preserves the value of the beverage within the bottle \cite{de2018forensics,ellis2017through,ellis2019rapid,kiefer2017analysis}. 

Raman spectroscopy is a non-destructive technique that measures the spectrum of light inelastically scattered by sample molecules, which has been used for both quantitative and qualitative analysis of spirits including whisky \cite{nordon2005comparison} and wines \cite{dos2018raman,mandrile2016controlling}. The technique is capable of analysing the concentrations of ethanol and methanol for quality control \cite{ashok2013optofluidic,kritzinger2025non} and, when the fluorescence spectrum of the sample is also analysed, can classify individual brands \cite{Ashok2011Nov,kiefer2017analysis}. Conventional Raman spectroscopy in a back-scattering configuration suffers from the simultaneous production of fluorescence and scatter from other materials present in the optical path, particularly originating from the glass of a bottle. This can confound and obscure the desired signal from the bottle contents. This problem can be overcome by a judicious choice of excitation wavelength \cite{ellis2019rapid}, by wavelength modulation techniques \cite{kritzinger2025non}, or by spatially offset Raman spectroscopy (SORS) and inverse-SORS (iSORS) \cite{nicolson2017through,ellis2017through}. In SORS, the illumination and collection points are laterally offset, ensuring that signal from the bottle surface is less likely to be collected than light which has laterally diffused within the sample. Rather than illuminating at a single point, iSORS uses an annular beam to illuminate with more power and thus increases signal intensity compared to SORS. Focus-matched iSORS (FM-iSORS), where a focusing annulus of light forms a ring on the bottle surface and a tight focus in the contents, has proven to be particularly well-suited to extinguishing the signal from glass bottles and has been used to identify whiskies, gins, vodkas and wines \cite{fleming2020through,Shillito2022focus,Kwanglee2024learning,kritzinger2025non,kritzinger2026through}.

Brandy is a spirit produced by distilling wine. Cognac is a prestigious variety of brandy, produced by the double distillation of white wine in copper pot stills, within the delimited Cognac region of France. The resulting colourless spirit, known as eau-de-vie, is aged in oak barrels and blended to produce the desired style of Cognac. The age designations, Very Special (VS), Very Superior Old Pale (VSOP), and Extra Old (XO), denote the minimum aging period of the youngest eau-de-vie in the blend, with VS requiring at least two years, VSOP a minimum of four years (though often aged longer), and XO demanding a minimum of ten years of maturation. As a high-value spirit, Cognac faces a significant challenge from counterfeiting, including fraudulent labelling and the use of non-compliant spirits \cite{sahakyan2023classification}.

Here, we show that through-bottle spectroscopy based on fluorescence and Raman scattering based on focus-matched iSORS can be used to identify the age classification and the bottling year of a bottle of Cognac, without opening the bottle or removing any sample. We also demonstrate that the approach can detect spoiling of Cognac that has been stored at higher temperatures than those recommended for maintaining product quality, and can be used for brand identification. Finally, we show that the approach can be used to separate genuine Cognacs from counterfeit products.

\section*{Results}
\subsection*{Focus-Matched iSOR spectroscopy}
\begin{figure} [b!]
\centering
\includegraphics[width=0.87\linewidth]{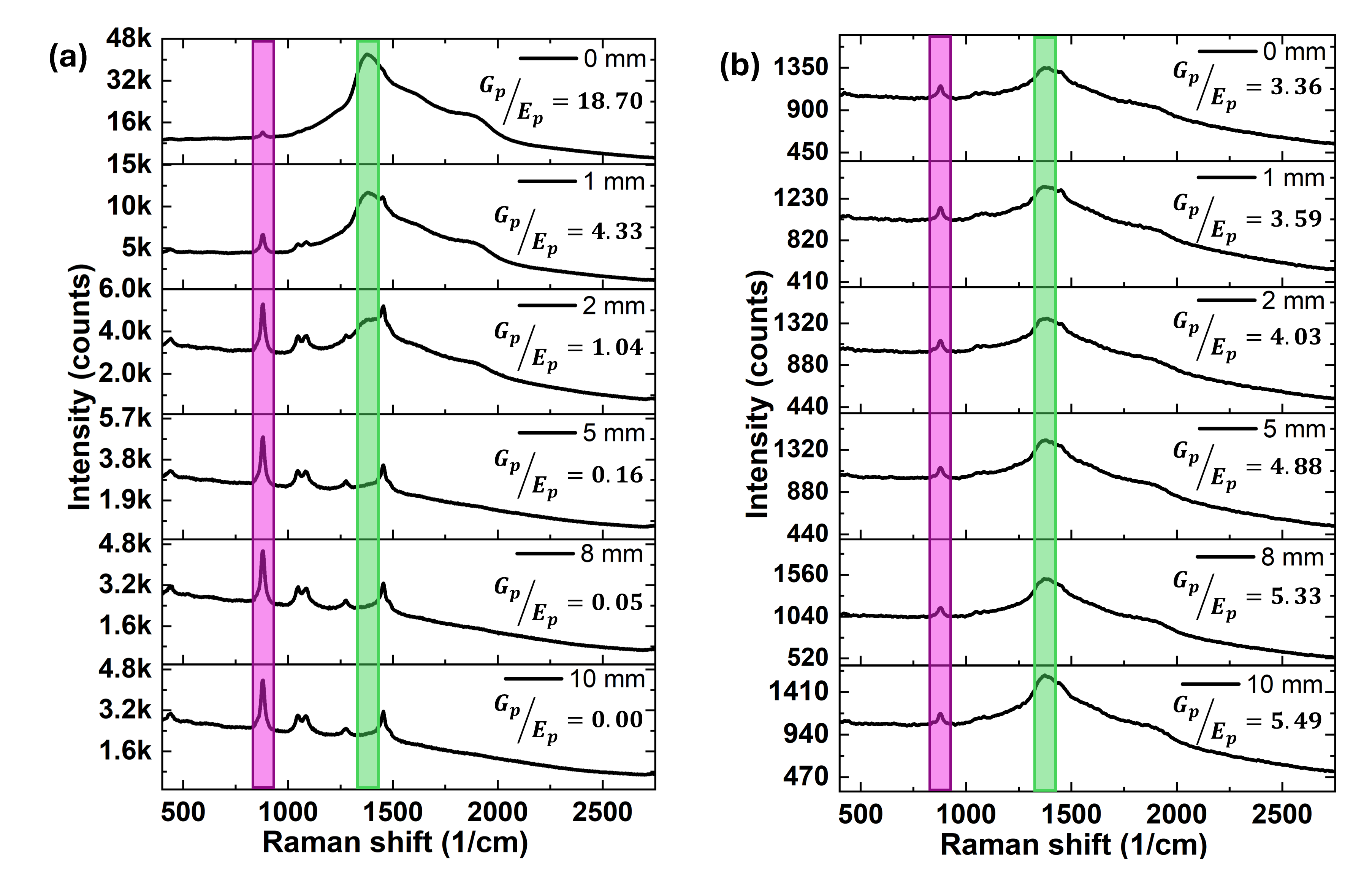}
\caption{\textbf{Comparison of the experimental Removal of Glass Signal between (a) FM-iSORS and (b) conventional iSORS.} Spectra obtained from a sealed bottle of Cognac, where each spectrum is acquired for a different bottle position. The legend for each panel shows the separation along the optical axis between the bottle surface, and the focal point of lens L$_{2}$ in Figure \ref{setup}. In FM-iSORS, increasing this separation has the effect of suppressing the signal due to glass which forms a peak centered at \mbox{1380 cm${^{-1}}$}. 
The Raman shift region corresponding to the dominant Raman peaks of ethanol and glass are highlighted in purple and green respectively. The ratio of glass-to-ethanol signal $G_{p}/E_{p}$ for FM-iSORS shows a decrease with increased focal distance, while the signal measured due to Raman scattering from the ethanol remains constant. In contrast, the ethanol peak is significantly weaker and does not substantially change with bottle position for iSORS, while the glass peak only reduces by $\sim$~15\% as the bottle moves away from the collection lens.}
\label{bottle_posi}
\end{figure}

The unique feature of the focus-matched iSORS is its ability to detect and optimize the content signal behind a barrier, while eliminating signal from the container. The method illuminates the bottle with a focusing ring of light, which forms its focus within the contents of the bottle. Back-scattered light is collected through the dark hole in the centre of the ring, to reduce signal from the container. This latter feature is shared with the more established iSORS technique \cite{Matousek2006, Yu2019}. However, the choice to focus the light rather than use a collimated ring, significantly increases the signal from the contents. This is illustrated in Figure \ref{bottle_posi} where a stack of spectra is plotted for both FM-iSORS and an iSORS configuration. 

In panel (a), each spectrum is acquired at a different bottle position relative to the focal point of excitation optics (lens $L_{2}$ in figure \ref{setup} in Methods), and the same positions are used in the iSORS configuration (panel (b)) for comparison. The bottle position was changed from the point when the laser is focused on the bottle (top stack - 0.0\,mm) to when the focus is 10\,mm inside the bottle (lowest for the FM-iSORS configuration). Due to the conical nature of the focusing beam, the size of the ring on the bottle increases for beams focused farther inside the bottle \cite{Shillito2022focus}. The largest ring size was \mbox{6\,mm}. For comparison purposes, the iSORS configuration was also set up with this ring size. Changing the bottle position does not alter the ring size in the iSORS configuration. These measurements were performed using a laser excitation wavelength of 785 nm, average laser power 250 mW, and 2 s integration time, five accumulations. 

The dominant Raman peak of ethanol (\mbox{880 cm${^{-1}}$}) and the glass (\mbox{1380 cm${^{-1}}$}) are highlighted in purple and green respectively. We use the height difference~(h$_d$) between the dominant ethanol peak and the fluorescence background (measured at \mbox{800 cm${^{-1}}$}) to indicate the strength of the Raman signal of ethanol. For the FM-iSORS configuration, h$_d$ was found to be approximately the same for all bottle positions, averaging \mbox{2068 $\pm$} 9 counts.  In contrast, the signal from the glass bottle decreases monotonically with increasing depth of focus. The glass peak is completely suppressed at a focus of 10.0\,mm inside the bottle. We use the height of the fluorescence at \mbox{1380 cm${^{-1}}$} to quantify the glass signal, using the spectrum obtained at the 10.0\,mm focal distance as a baseline value. Therefore, the glass-to-ethanol ratio, G$_p$/E$_p$, is calculated as [(glass~peak$_{1380}$)$_i$-(glass peak$_{1380}$)$_{10.0\,mm}$]/h$_{di}$ where $i$ represents each spectrum. When the bottle is placed in the beam focus, G$_p$/E$_p$ was 18.70. As the bottle was moved, this falls monotonically until it was completely insignificant at depths beyond 10.0\,mm.  Therefore, for all subsequent measurements with FM-iSORS, all bottles were placed to ensure that the depth of focus was at least 10.0\,mm inside the bottle. 

For the conventional iSORS geometry, h$_d$ was found to be a factor of 18 smaller (\mbox{114 $\pm$ 1} counts) for the same power and integration parameters, and was independent of bottle position. Moreover, G$_p$/E$_p$ is always larger than unity, and slowly increases as the bottle approaches L$_{2}$ due to a slight increase in the collection of this light into the spectrometer. This is due to the close proximity of the glass bottle to the focal plane of the lens. The peak heights of the dominant ethanol peak and the glass-to-ethanol ratios relative to the positions of the bottle are shown in supplementary section S1.

These measurements demonstrate that FM-iSORS can eliminate signals from the containers without significant losses in the strength of the content response. Furthermore, our data confirm that the focusing of the excitation beam in FM-iSORS leads to a higher signal from the contents than is achieved in conventional iSORS for comparable parameters. This increased signal allows FM-iSORS to operate at reduced power: in the remainder of this work, we use only \mbox{100 mW} of excitation power.

\subsection*{Age classification and batch identification of Cognac}

\begin{figure} [ht!]
\centering
\includegraphics[width=0.82\linewidth]{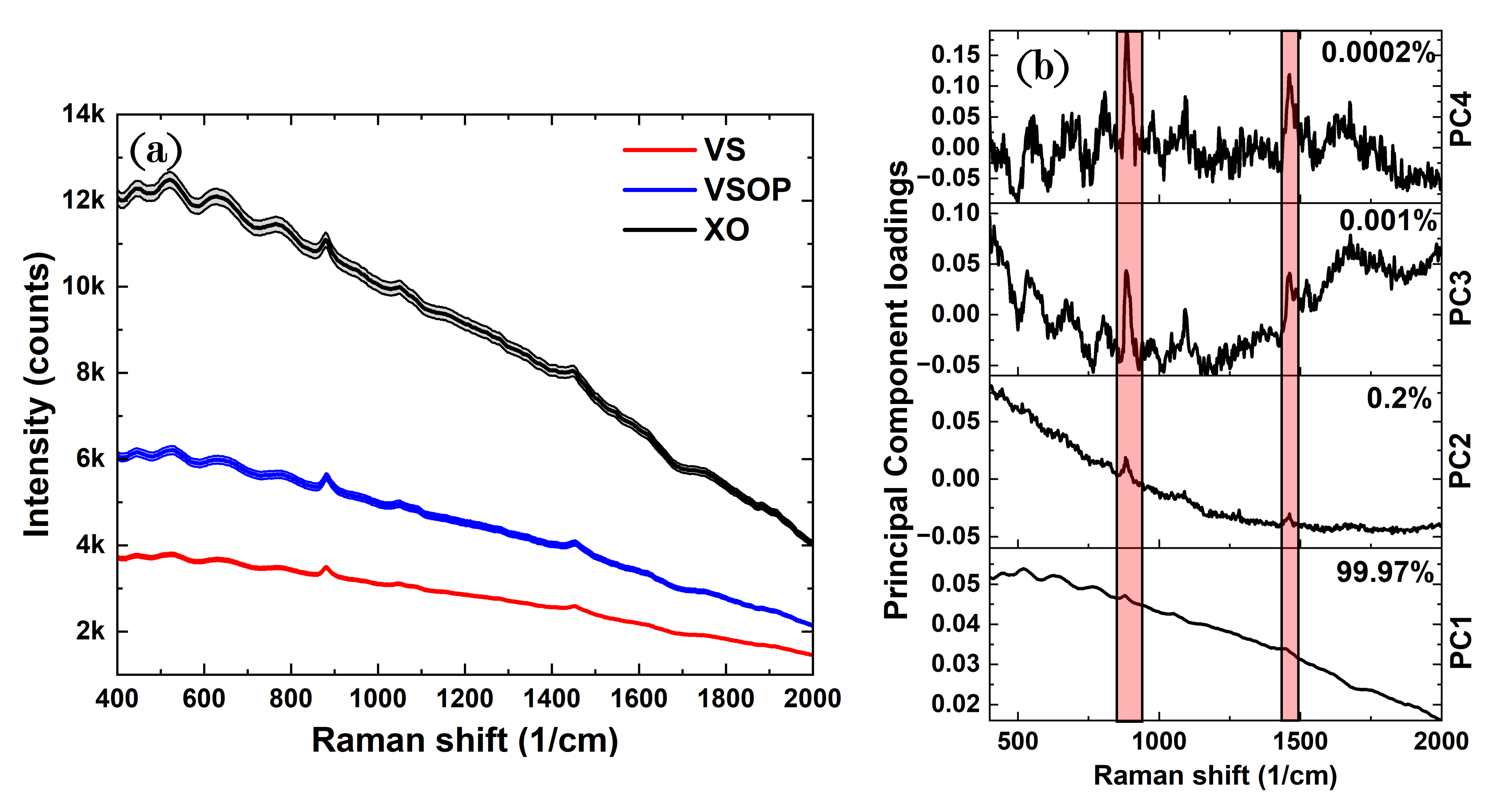}
\caption{\textbf{Spectral analysis of VS, VSOP and XO age classifications from a single producer.} (a) The average spectrum (solid line) and standard error (shaded area) of each age classification shows visible differences in the broad fluorescence profile. (b) Principal component loadings of the first four principal components of the spectra. The variance captured by each principal component is included within the relevant panel. Red shaded regions show the positions of the two dominant Raman peaks of ethanol at \mbox{880 cm$^{-1}$} and at \mbox{1460 cm $^{-1}$}.}
\label{fig2:brandy_types}
\end{figure}

Using focus-matched iSORS, we performed an initial spectroscopic analysis of different age classifications of Cognac from a single supplier, where the Cognac remained sealed in its original bottle. All brandy and Cognac samples used in this investigation consist of 40\% ethanol, with a mixture of other organic compounds and water. The organic compounds, collectively referred to as congeners, are made up of esters, organic acids, aldehydes, and higher-order alcohols. These congeners, together with the extracts from the wooden cask and the colouring material, including caramel, contribute to the distinct flavour and aroma of the specific product \cite{Stanzer2023May_congeners}. Upon excitation, the molecules undergo both fluorescence and Raman scattering, as can be seen from the average spectra obtained for each age classification of Cognac (VS, VSOP, and XO) in Figure \ref{fig2:brandy_types}a. As with other spirits, the broad peak is a result of fluorescence from the congeners, while the narrow features (e.g. at \mbox{880 cm$^{-1}$} and at \mbox{1460 cm $^{-1}$}) are Raman peaks from the ethanol \cite{Ashok2011Nov}. These peaks represent the C-C-O symmetric stretching vibrations and the \text{CH}$_3$ antisymmetric deformation modes respectively \cite{Emin}. An increase in fluorescence was observed to be correlated with the darker appearance of the longer-aged Cognacs. Despite the change in the fluorescence background, the Raman peaks of the dominant ethanol content remain the same.  The Raman contributions from the congeners were not detected over the high fluorescence signal. Unlike other Raman analyses where the fluorescence background is removed, the analysis carried out here included the fluorescence as it has been shown to contribute to the successful classification of liquor \cite{Ashok2011Nov}.

\begin{figure}[ht!]
\centering
\includegraphics[width=0.82\linewidth]{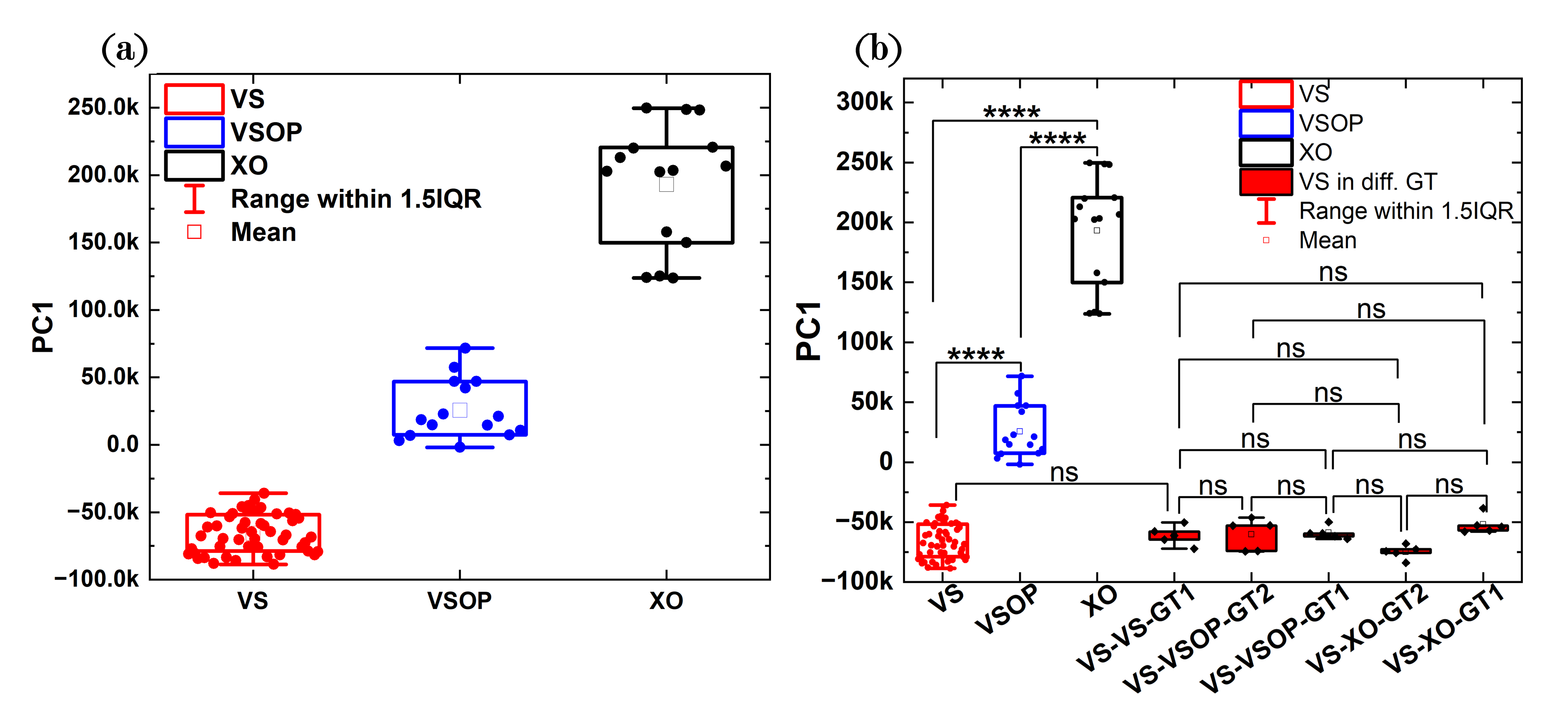}
\caption{\textbf{Identification of age classification is unaffected by bottle substitution.} In (a), we present a box plot of PC1 values for the different age classifications of Cognac (VS, VSOP and XO), showing how they can be discriminated by fluorescence alone. Each point represents the PC1 score for a single measurement. In (b), we show box plots comparing the PC1 scores obtained from the original test blends with other bottles refilled with VS blend. Data are presented as PC1 score values. Red data and boxes denote low end VS bottles while blue and black boxes represent VSOP and XO respectively. Each box represents the 25\% to 75\% range. Solid red boxes represent data from substituted bottles of different glass types (GT). Data were analysed with PCA and one-way ANOVA with Tukey's multiple comparison test. Asterisks indicate statistical significance between treatment groups. ****P<0.0001 ns: no significance.}
\label{box_all_gen_bottles}
\end{figure}

A multivariate principal component analysis (PCA) of all spectra recorded from all bottles identifies four significant principal components of the spectra, the loadings of which are shown in Figure \ref{fig2:brandy_types}b. One can observe that the first principal component (PC1) dominates with a 99.97\% variance. The main contributing feature of PC1 is the broad fluorescence profile. Higher PCs show decreasing contributions from the fluorescence spectrum and increasing contributions from the Raman spectrum of the ethanol. The C-O stretching vibration, which corresponds to a Raman peak at \mbox{1050 cm$^{-1}$}, is more visible in PC3 and PC4 than in the original spectra.
Figure \ref{box_all_gen_bottles}(a) displays a box plot of the PC1 scores obtained by projecting each individual spectrum onto the PC1 eigenvector. A clear separation between the spectra of the three Cognac age classifications is observed. This single parameter is sufficient to uniquely identify the age classification of the Cognac, as is demonstrated by the results of leave-one-out cross-validation (LOOCV) shown in Table \ref{table_bottles}. A 98\% prediction accuracy was obtained from 50 measurements of VS bottles and a 100\% accuracy was obtained for the other two age classifications.

\begin{table}[ht!]
\centering
\renewcommand{\arraystretch}{2}
\caption{LOOCV table for the 3 main Cognac age classifications. }
\vspace{0.1em}
\label{table_bottles}
\begin{tabular}{|c|c|c|c|c|c|c|}
\cline{3-6}
\multicolumn{2}{c|}{} & \multicolumn{4}{c|}{\makecell{Identified Cognac}} & \multicolumn{1}{c}{} \\ \cline{3-7}
\multicolumn{2}{c|}{} & \makecell[c]{VS \rule[-8pt]{0pt}{20pt}} & \makecell[c]{VSOP \rule[-8pt]{0pt}{20pt}} & \makecell[c]{XO \rule[-8pt]{0pt}{20pt}} & \makecell[c]{Unknown \rule[-8pt]{0pt}{20pt}} & \makecell[c]{\% accuracy \rule[-8pt]{0pt}{20pt}} \\ \hline
\multirow{3}{*}{\rotatebox{90}{\makecell{Actual Cognac}}} & \makecell[l]{VS \rule[-8pt]{0pt}{20pt}} & 49 & 1 & 0 & 0 & 98\\ \cline{2-7}
& \makecell[l]{VSOP \rule[-8pt]{0pt}{20pt}} & 0 & 15 & 0 & 0 & 100 \\ \cline{2-7} 
& \makecell[l]{XO \rule[-8pt]{0pt}{20pt}} & 0 & 0 & 15 & 0 & 100 \\ \hline
\end{tabular}
\end{table}

One form of fraud seen with spirits is bottle substitution - for example, replacing an expensive higher age classification Cognac with a less expensive lower age classification. To determine whether bottle substitution could cause misidentification of the Cognac, we poured VS Cognac into different bottles used by the same producer. The bottles were manufactured by different commercial glassmakers in two different locations with differing chemical compositions. We refer to these compositions as GT1 and GT2. Bottles of both glass types with different shapes (associated with their intended contents' age classification) were used. A separate analysis of the bottles and the glass types can be found in supplementary fig. S2. Five spectra consisting of five averages per spectra were recorded, and projected against the same PC1 eigenvector shown in Figure \ref{fig2:brandy_types}b. Figure \ref{box_all_gen_bottles}(b) shows the box plot of the different bottle substitutions. It is evident that irrespective of the glass type or the age classification of the bottle used, the PC1 score ranged between the high and low values of the unopened VS in its original bottle. To further confirm this, we performed a one-way ANOVA using the Tukey's multiple comparison test \cite{Tukey} and a 95\% confidence interval difference. From Figure \ref{box_all_gen_bottles}(b), it is observed that there is greater statistical significance between the different original test blends with P< 0.0001. However, there was no statistical significance (ns) between the VS in its original state and the substituted bottles as P-values ranged between 0.72 and 0.99. Thus, a lower age classification Cognac can clearly be distinguished from a higher age classification product, regardless of bottle substitution. This is because the spectroscopy configuration used here is particularly designed to avoid the collection of any signal from the glass walls of the bottles (see Methods) \cite{fleming2020through, Shillito2022focus, Kwanglee2024learning}. 

\begin{figure} [ht!]
\centering
\includegraphics[width=0.8\linewidth]{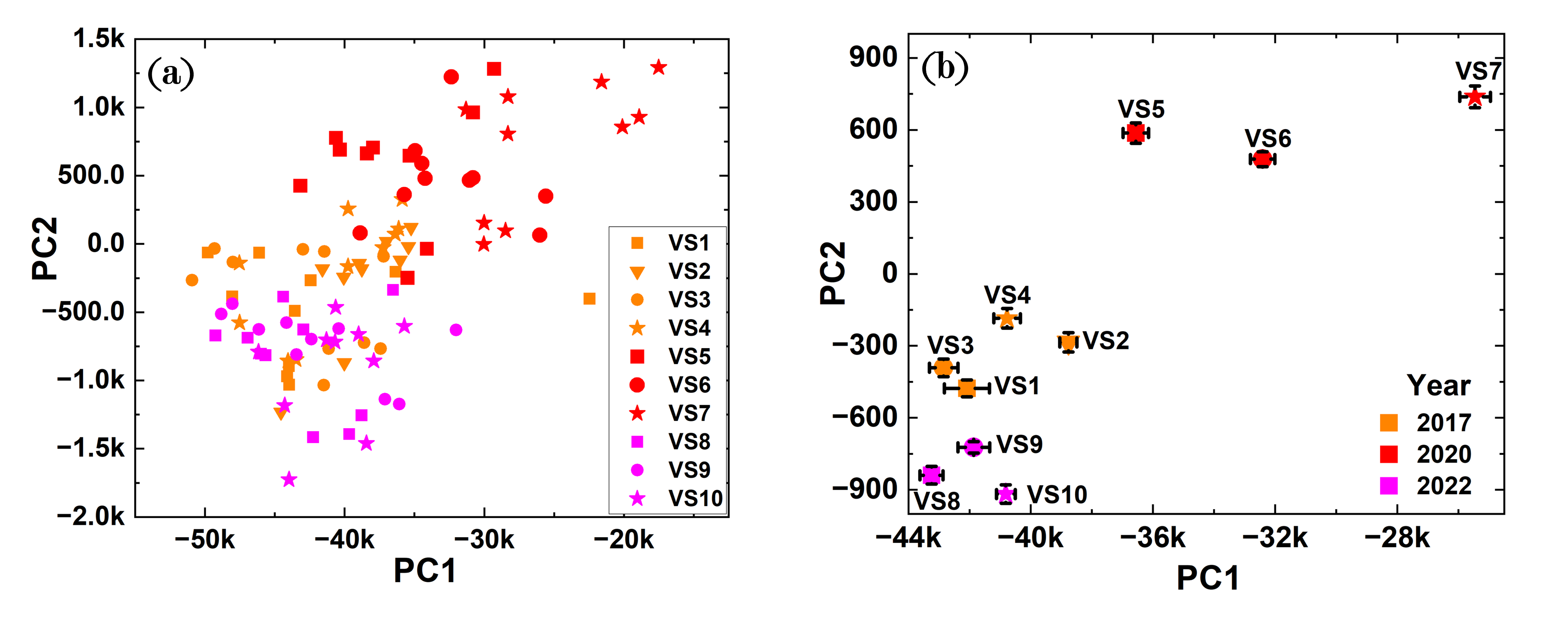}
\caption{\textbf{Clustering of spectra from different production years.} Principal component score plots for VS bottles show clustering based on the year of production. (a) PC1-PC2 scatter plots for individual VS bottle measurements shows overlapping clusters. (b) However, averaged measurements of each bottle shows separation by year along the PC2 axis. Data in (b) are presented as mean $\pm$ SEM. Colours denote the year of production, while symbol shapes denote different bottles.}
\label{fig5:VS_PCA_and_ave}
\end{figure}

So far, we have used only the first principal component of the eigenbasis identified by PCA trained on bottles from three age classifications. While this eigenvector allowed us to discriminate age classification, there is further information encoded in the other PCs. Plotting the value of PC1 vs the value of PC2 for only VS bottles in Figure \ref{fig5:VS_PCA_and_ave}a, we observe that there is no significant separation along the PC1 axis, as expected. By plotting individual spectra, we see overlapped clusters that indicate the year of production (taken from the bottle label). However, on averaging all acquisitions recorded for each bottle, we observe a clear clustering along the PC2 axis which indicates the year of production, as shown in Figure \ref{fig5:VS_PCA_and_ave}b.

\subsection*{Detection of temperature-dependent changes in Cognac}

Temperature is known to play a significant role in the development of spirits, with elevated temperature degrading the sensory attributes and shelf-life of the liquor \cite{Ferreira2022Jan_temp,Scrimgeour2015Dec_temp,Sivertsen2001}, and is the main parameter contributing to major changes in the chemical profile of beer \cite{Scrimgeour2015Dec_temp}. While brandy and other spirits are generally more stable than wine due to higher alcohol content, they are still susceptible to temperature-induced changes \cite{temp_cognac}. For example, high temperatures exacerbate the loss of volatile compounds \cite{Hopfer2012, PerezCoello2003}. Fluctuating temperatures can lead to expansion and contraction of the liquid, potentially compromising the integrity of the seal and allowing more air (oxygen) into the bottle \cite{Scrimgeour2015Dec_temp}, accelerating oxidation reactions and altering acidity, flavour and aroma \cite{Cutzach1999, BENUCCI2020108732}.

\begin{figure} [t!]
\centering
\includegraphics[width=0.8\linewidth]{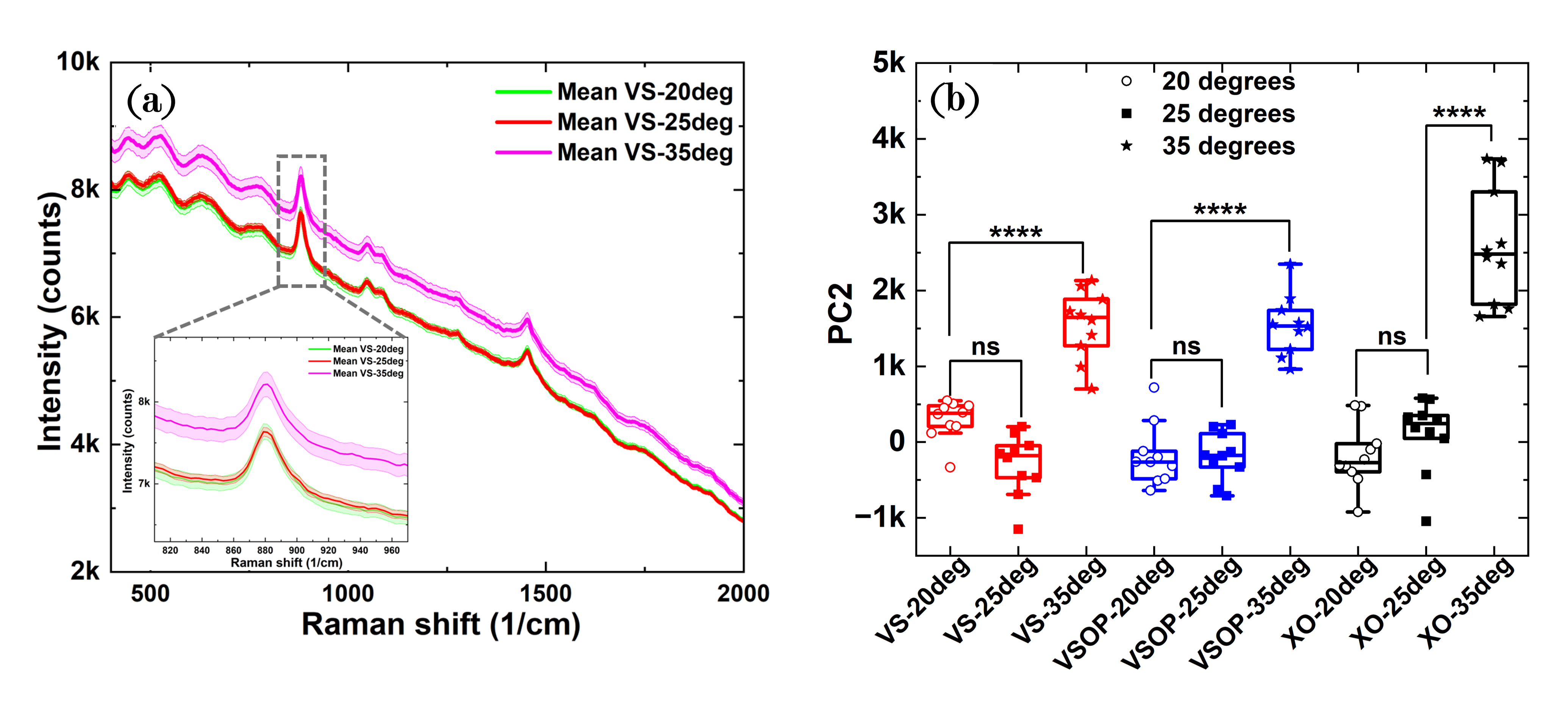}
\caption{\textbf{Cognac stored at higher temperatures shows spectral changes.} (a) Through-bottle spectra of VS and (b) principal component analysis for VS, VSOP and XO Cognac stored at different temperatures: 20\textdegree C (green), 25\textdegree C (red) and 35\textdegree C (magenta). All data were recorded with 100 mW power and 2s integration time. Inset in (a) shows the region around the dominant \mbox{880 cm${^{-1}}$} ethanol peak highlighting the strong overlap between samples stored at acceptable temperatures. In (b), VS, VSOP and XO are represented in red, blue and black respectively. Data were analysed with PCA and one-way ANOVA with Tukey's multiple comparison test. Asterisks indicate statistical significance between treatment groups. ****P<0.0001 ns: no significance. Cognac stored at 35\textdegree C is consistently shown to have a statistically significant shift in the PC2 value.}
\label{fig6:brandy_types_temp}
\end{figure}

We assess the potential use of through-bottle spectroscopy for quality control by analysing bottles of Cognac of each age classification which had been stored in bottles at three different temperatures (20\textdegree C, 25\textdegree C and 35\textdegree C). Each bottle was measured in 5 different orientations (i.e. bottle rotated to 5 different positions with no label obscuring the beam path) with the spectra constructed from 5 averages at each orientation. Figure \ref{fig6:brandy_types_temp}a shows the spectra of the bottles filled with VS stored at the different temperatures. The plots for VSOP and XO bottles are presented in the supplementary materials (S3). The effect of storage at 35\textdegree C can be seen in all age classifications: there is an overlap in the spectra for the bottles stored at 20 and 25\textdegree C, while the bottles stored at 35\textdegree C show a higher level of fluorescence.

A box plot comparison of the PC2 scores for the different age classifications stored at different temperatures is shown in Figure \ref{fig6:brandy_types_temp}b. For all blends, bottles stored at 35\textdegree C show an offset along the PC2-axis while bottles stored at 20 and 25\textdegree C overlapped. A one-way ANOVA show there is no statistical significance between the data for bottles stored at 20 and 25\textdegree C while presenting a 4-star (P<0.0001) significance between those bottles and those stored at 35\textdegree C. A detailed sheet of the ANOVA analysis for all data points is shown in table S1 in the supplementary materials. This result highlights the potential applicability of through-bottle spectroscopy for quality control in Cognacs and brandies.

\subsection*{Cognac and brandy brand identification}
\begin{figure} [ht!]
\centering
\includegraphics[width=0.75\linewidth]{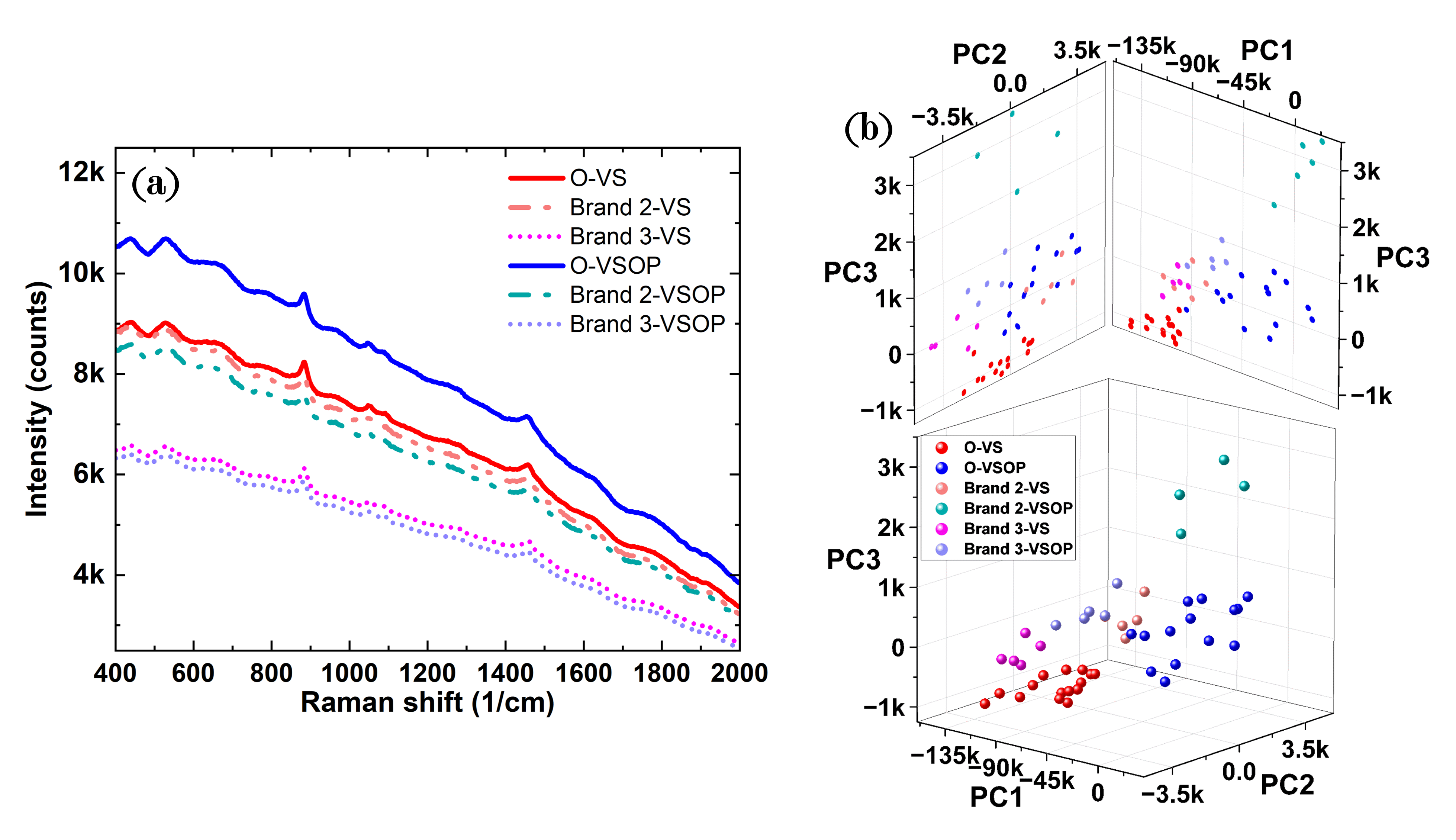}
\caption{\textbf{Spectral clustering of different Cognac brands.} (a) Raman spectra and (b) principal component analysis of different VS (reds) and VSOP Cognacs (blues) in sealed original bottles. }
\label{fig7:other brandies}
\end{figure}

For a multi-billion dollar industry \cite{TSB1, GVR1}, it is essential for each producer to be distinguishable by the uniqueness of their products. In this section, we investigate the spectral separation of Cognacs from three different producers. Figure \ref{fig7:other brandies}a shows the average spectra of the previously analysed Cognacs (termed as original and referenced with (O-)) and the spectra from other producers, all in sealed original bottles. All VS bottles are in shades of red while VSOP are in shades of blue. It can be seen that the fluorescence levels of all three brands are comparable. 

A principal component analysis of these spectra was performed, and the first three principal component scores are plotted in Figure \ref{fig7:other brandies}b. The top panel shows the 2D projections with the full 3D projection shown below. No 2D projection is sufficient to fully separate the different Cognacs: there is an overlap in the scores of VS from brand-2 with VS from brand-3 in the PC1-PC3 plot and VSOP (O-VSOP) in the PC2-PC3 plots. However, there is a clear separation among these brands when the full 3D space is considered. This indicates that despite the closeness in spectral properties and fluorescence among brands, a clear classification is possible when a volumetric projection is performed.

\subsection*{Counterfeit detection}
A major concern for producers is product counterfeiting, which can cause financial losses and reputational damage. Counterfeit samples were prepared and placed in original bottles to mimic the original Cognacs of interest. More specifically, the counterfeits consisted of a mixture of other spirits with colourants, so that a colour-match to the original Cognac was achieved. 
\begin{figure} [ht!]
\centering
\includegraphics[width=0.7\linewidth]{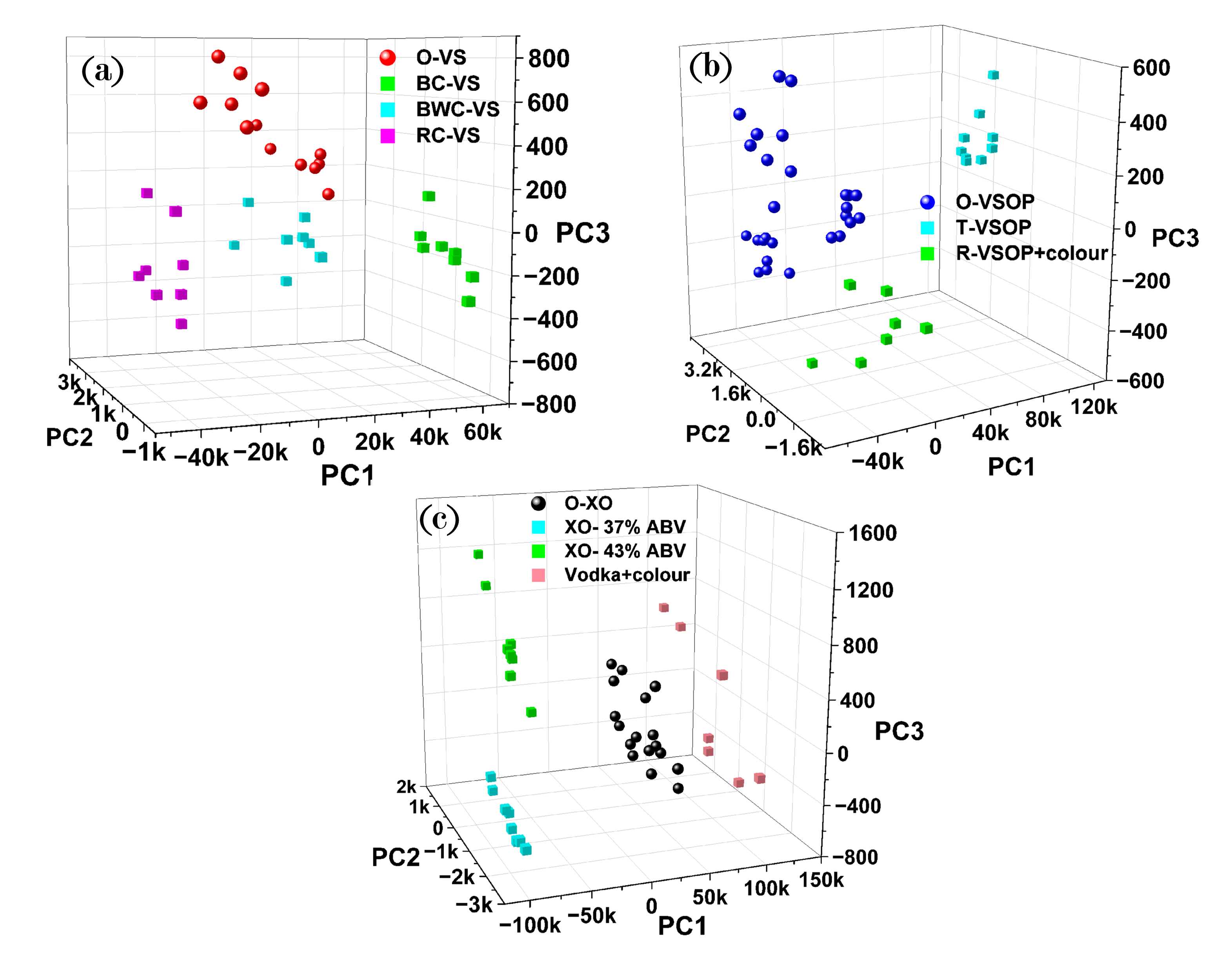}
\caption{\textbf{Spectral clustering of genuine and counterfeit products.} 3D principal component space of original (O-) Cognacs with counterfeits packaged in the bottles of the associated original product. Counterfeits ranged from mixtures of brandy with colour (BC) to Vodka with colour. A complete compositional list of all counterfeits is given in the supplementary materials.}
\label{PCA_counterfeits}
\end{figure}

 Figure \ref{PCA_counterfeits} shows the 3D PC space of the various counterfeits compared with the original Cognacs. To streamline classification, fake bottles were only compared with the originals matching the bottles in which they came. Figure \ref{PCA_counterfeits}a is a representation of the Original VS with all the counterfeits provided in VS bottles. A brandy with colour (BC-VS), blended whisky plus colour (BWC-VS) and rum with colour (RC-VS) were investigated. Similarly, the counterfeit VSOP bottles were a commercial brandy and a mixture of different of rum and colour (Figure \ref{PCA_counterfeits}b) while the fake XO bottles had different alcohol concentrations and a mixture of vodka and colour (Figure \ref{PCA_counterfeits}c). A probabilistic assessment of the contents of each bottle was performed by LOOCV, the results of which are detailed in a confusion matrix in Table \ref{probtable}. As expected 
 from the results in Section 3.1, there is no misidentification crossing age classifications. For samples in VS bottles, the through-bottle spectra give 91\% true-positives and 9\% false-negatives identified as counterfeits. Through-bottle spectra of samples in VSOP and XO bottles both achieved 96\% true-positive rates, while the true-positive rate in identifying counterfeits was 98\% with a low false-positive rate of 2\%. A detailed prediction for each individual bottle is shown in table S2 (in supplementary materials).  

\begin{table}[t!]
\centering
\caption{Probabilistic assessment of genuine and counterfeit Cognacs} \label{probtable}
\begin{tabular}{|l|c|c|c|c|}
\cline{2-5}
\multicolumn{1}{c|}{} & VS & VSOP & XO & Other \\ \hline
Genuine VS & 91\% & 0\% & 0\% & 9\% \\ \hline
Genuine VSOP & 0\% & 96\% & 0\% & 4\% \\ \hline
Genuine XO & 0\% & 0\% & 96\% & 4\% \\ \hline
Other & 2\% & 0\% & 0\% & 98\% \\ \hline
\end{tabular}
\end{table}

\section*{Discussion}
In this work, we have shown that the three age classifications (VS, VSOP and XO) of Cognac can be identified using through-bottle spectroscopy and principal component analysis, with an accuracy of 98\%. The analysis was shown to be independent of the glass type and shape of the bottle, which is important in an anti-counterfeiting scenario as it enables identification of substitutions of lower value product into more expensive bottles. Moreover, the through-bottle approach was also shown to achieve separation of the spectra obtained from different brands of VS and VSOP Cognacs and indicates an ability to identify the year of production. A probabilistic assessment of genuine and counterfeit samples using leave-one-out cross-validation achieved a sensitivity of 94\% in identifying genuine products and correctly identifies 98\% of counterfeit product. 

From a quality control perspective, the methods shown are also capable of discriminating Cognacs of all three age classifications that have been stored at 35$^{o}$C from those stored at or below 25$^{o}$C. The detected change in fluorescence is likely due to increased oxidation and breakdown of esters within the spirit. A detailed, temperature- and time-resolved co-measurement of the chemical profile using liquid chromatography or similar approaches was not possible within our facilities. In future, such additional measurements could provide interesting correlative data to understand the changes in the fluorescent profile caused by the increased temperature. Taken in combination, these results strongly suggest that through-bottle spectroscopy using a focus-matched offset geometry is well suited to quality control and anti-counterfeiting assessment of brandy and Cognac.

\section*{Methods}
\subsection*{Spectroscopic system}

\begin{figure} [ht!]
\centering
\includegraphics[width=0.76\linewidth]{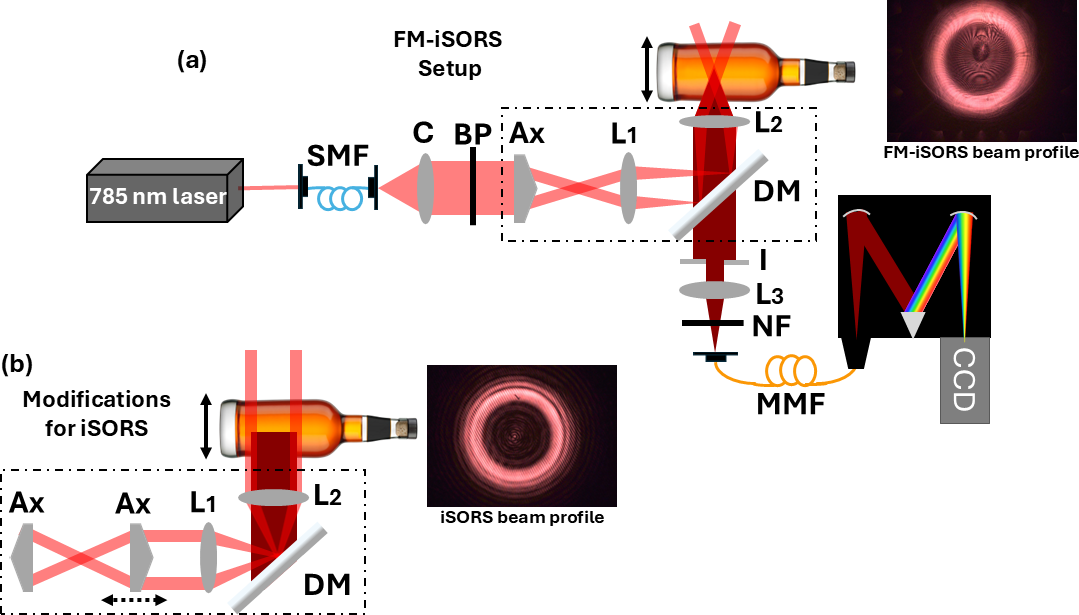}
\caption{ \textbf{Experimental geometries of the focus-matched iSORS for through-bottle detection (a) and conventional iSORS detection (b).} Both geometries were detected in a back-scattering configuration. The beam profiles for the respective configurations are shown next to the bottles. SMF: single-mode fibre; C: collimating lens; Ax: axicon lens; BP: laser line filter; DM: dichroic mirror; I: iris; NF: notch filter; MMF: multi-mode fibre (200 \textmu m core diameter); L$_{1}$-L$_3$: Lenses}
\label{setup}
\end{figure}

We employed the focus-matched inverse spatially-offset Raman (FM-iSORS) technique. The Raman measurements were performed using a free-space system, consisting of a Spectra-Physics 3900s Ti:Sapphire tunable laser for excitation and an Andor Shamrock SR-303i spectrometer for spectral measurement. The experimental setup in this section was based on that previously demonstrated by Fleming et al. \cite{fleming2020through} and Shillito et al.\cite{Shillito2022focus}, and is shown in Figure \ref{setup} (a). In brief, laser light with an excitation wavelength of \mbox{785 nm} was sent through a single mode fibre (SMF) with the output beam collimated using a 30X plano-convex aspheric objective lens (C) (Newport 5723-B-H) and line filtered (Semrock LL01-785). The beam passed through an axicon lens (Ax, $\alpha$=10\textdegree, Thorlabs AX255-B), producing a Bessel beam for which the Fourier transform realized in the far-field is an annular beam. Achromatic lenses L$_1$ \mbox{(f= 100\,mm)} and L$_2$ \mbox{(f= 40\,mm)} in a 4-f geometry were used to relay the Bessel beam into the bottle, while forming a ring illumination on the surface. A dichroic mirror (DM) (Semrock LPD02-785RU) separated the excitation and back-scattered light. Direct back-scatter from the bottle surface was blocked with an iris (I). Rayleigh scatter contributions to the back-scatter were removed by a notch filter (NF) (Semrock NF03-785E). The back-scatter was imaged onto the collection tip of a 200\,\textmu m-core multimode fiber (MMF) using L$_{2}$ and lens L$_3$ \mbox{(f= 50\,mm)} in the 4-f configuration. The collected signal was analyzed using a CCD spectrometer (Andor Shamrock SR-303i) with a 0.1 nm resolution. To convert the FM-iSORS system into a conventional iSORS geometry, the boxed part in the FM-iSORS setup was reconfigured into the part shown in panel (b) in Figure \ref{setup} with the addition of a second axicon (Ax, $\alpha$=10\textdegree, Thorlabs AX255-B) and replacing L$_1$ with a \mbox{f= 40\,mm} lens to create a 4f-configuration with L$_2$. The separation distance between the 2 axicons in the iSORS configuration controls the diameter of the collimated annular ring. The beam profiles for both configurations are shown next to their respective panels. The ring in the FM-iSORS geometry has a brighter core due to the focusing effect of the lens L$_2$ while that of the iSORS is seen to have a darker core due to the collimation.

All spectra were collected in the spectral range \mbox{140-2700 cm${^{-1}}$} with a laser power of 100 mW in the bottle, an integration time of 2 seconds and 5 accumulations. The system was calibrated daily using paracetamol with the SNR quantified by comparing the height of the largest paracetamol peak at \mbox{858 cm${^{-1}}$} to the standard deviation $\sigma$ of the noise. The value of $\sigma$ was determined by taking the standard deviation of the counts in the \text{``silent''} region where paracetamol peaks were absent (\mbox{2400-2700 cm${^{-1}}$}). For the Cognac measurements, each bottled sample had 5 replicate spectra collected, which were preprocessed with SpectraGryph (v1.2 16.1) to crop to the \mbox{400-2000 cm${^{-1}}$} range and remove spikes. To keep the fluorescence profile of the bottle content, no baseline substraction was performed. All bottles were measured under the same conditions. All analyses of bottles of altered or substituted Cognac were performed double-blind.

In assessing the age classification and batch discrimination capabilities of the approach, a total of 16 samples of Cognac in their original bottles (comprising 10 bottles of `Very Special'(VS) 2 year old Cognac, 3 `Very Superior Old Pale' (VSOP) 4 year old Cognac, and 3 `Extra Old' (XO) 10 year old Cognac) from a single brand were measured 10 times each. For the purpose of this work, these samples will be simply referred to as VS, VSOP, and XO, unless they are compared to products from other manufacturers in which case they will be prefixed with the original (O-). The VS and VSOP bottles consisted of 3 batches each (2017, 2020 and 2022 for VS and 2015, 2011 and 2019 for VSOP) while the XO was from 2016.  For storage temperature dependency measurements, the producer provided nine identical bottles containing VS, VSOP or XO stored in a temperature-controlled cabinet for 2 years at rest temperatures of 20\textdegree C, 25\textdegree C, or 35\textdegree C.

For brand identification tests, spectra of VS and VSOP Cognacs from three different producers were compared. For counterfeit identification tests, mixtures of different spirits and colourant (caramel colouring agent E150A) were produced in original bottles from one producer. Colourant was added to obtain a visual colour match to original product in an identical bottle. The `counterfeits' packaged in VS bottles were brandy, blended whisky, and rum, each coloured to match the original VS Cognac. A rum with added colourant, and a commercial brandy which already appeared to colour-match the original VSOP Cognac, were prepared in VSOP bottles. A vodka with colourant was prepared in an XO bottle, as were original XO Cognac diluted with water to achieve 37\% alcohol by volume (ABV) and original XO Cognac with added ethanol to achieve 43\% ABV.

\subsection*{Principal component analysis and leave-one-out cross-validation prediction}
To analyse the Raman spectra, PCA was used to find the vector space in which all spectra are maximally separated. The eigenvectors of this analysis were inspected for qualitative understanding of the defining features. When the spectra are plotted in the PC space, spectra from similar samples cluster to the same position. To quantitatively assess the ability of our system to identify a sample based on a single spectrum, we designed an algorithm which projects the spectrum of an unknown sample into the PC space of known samples, and uses this to give a likelihood that the measured sample belongs to any of the clusters. 

To do this, a scoring algorithm produces a confidence score for each known sample, on the statement that the unknown sample is the known sample. This confidence score is valued between 0 and 1: the greater the value, the greater the confidence. These scores are calculated by first generating an intermediate confidence score ICS$_{ij}$ for all principal axes and samples:

\begin{equation}
\label{equ:IntermediateCS}
ICS_{ij} = \exp{\left(\frac{-{\left(u_{i} - \bar{u}_{ij}  \right)}^{2}}{2\sigma_{ij}}\right)}
\end{equation}
where $u_{i}$ is the position of the unknown sample in the i\textsuperscript{th} dimension of the PC space, $\bar{u}_{ij}$ is the mean score of the j\textsuperscript{th} known sample in the i\textsuperscript{th} dimension of the PC space, and $\sigma_{ij}$ is the standard deviation score of the j\textsuperscript{th} known sample in the i\textsuperscript{th} dimension of the PC space. To find the confidence score corresponding to each sample, the minimum of $ICS_{ij}$ along all dimensions of the PC space is found. To make the prediction, a final confidence score is produced by finding the maximum of the minimum scores of each known sample. The whole calculation of this final score is summarized by:

\begin{equation}\label{equ:FinalCS}
\text{Final Confidence Score} = \max_{j=1, \dots, N_{j}}\left(\min_{i=1, \dots, N_{i}}\left({ICS_{ij}}\right)\right),
\end{equation}
where $N_{j}$ is the number of known samples, and $N_{i}$ is the number of dimensions in the PC space. If this final confidence score is below a user defined threshold the unknown sample is predicted as ``not a known sample''. Otherwise, the algorithm identifies the unknown sample as the closest sample in the training space.

\section*{Acknowledgments}
We thank Dr Mingzhou Chen, Dr Georgina Shillito, Rory Buchanan and Rachel Barron for early inputs to apparatus design. We acknowledge funding from University of St Andrews Impact \& Innovation Fund, EPSRC Impact Accelerator EP/X525819/1, EPSRC Place Based Impact Acceleration Account EP/Y024109/1, and ARC Laureate Fellowship (grant FL210100099).

\section*{Data Availability}
Underpinning data will be made available via the University of St Andrews Research Repository.

\bibliography{sample}



\end{document}